\documentstyle[12pt,epsfig]{article}

\begin{document}

 \setlength{\baselineskip}{16.0pt}
\begin{center}{\bf  A numerical study of the spectrum and eigenfunctions on a
tubular arc} \vskip 6pt \centerline{ Lonnie Mott, Mario Encinosa,
and Babak Etemadi} \vskip 2pt \centerline{Department of Physics}
\vskip 2pt \centerline{Florida A \& M University}\vskip 2pt
\centerline { Tallahassee, Florida 32307} \end{center}
 \setlength{\baselineskip}{16.0pt}
\begin{abstract} \noindent The Hamiltonian for a particle
constrained to move on the surface of a curved nanotube is derived
using the methods of differential forms. A two-dimensional
Gram-Schmidt orthonormalization procedure is employed to calculate
basis functions for determining the eigenvalues and eigenstates of a
tubular arc  (a nanotube in the shape of a hyperbolic cosine) with
several hundred scattering centers. The curvature of the tube is
shown to induce bound states that are dependent on the curvature
parameters and bend location of the tube.

\vskip 8pt \noindent PACS number(s): 03.65.Ge, 68.65.-k

\end{abstract}
 \vskip6pt
\noindent $\bf{1. \ Introduction}$ \vskip 6pt The quantum mechanics
of reduced dimensionality systems [1-7] has become a subject of
importance in modelling devices for which the geometric properties
of the device become a factor in influencing its behavior [8-15].
Carbon nanotubes are such objects; the electronic properties of
straight nanotubes are determined primarily by the chirality of the
tubes and are well understood [16]. However, bent, curved, and
toroidal CNT's have been observed [17], and their potential
application as device elements makes it necessary to model such
structures as well.

\vskip 6pt In this paper we perform a curved manifold Schrodinger
equation calculation to determine the spectrum and eigenfunctions
for a particle constrained to move on the surface of a curved
nanotube with delta function site potentials. The methods of
differential forms are used to derive the Hamiltonian of a
particle constrained to move on the surface of a curved nanotube.
Initially the particle is taken to be in three-dimensional space;
it is then confined to move on a two-dimensional curved manifold
by a potential everywhere normal to the surface. This reduction in
dimensionality yields a curvature-dependent potential $V_D$
[1-7,18-20] that is added to the Hamiltonian. Then, using a
hyperbolic cosine shape function, the eigenvalues and eigenvectors
for the low-lying states of a tubular arc with and without
$\delta$-function potentials are calculated with a basis set
expansion wherein  a two-dimensional Gram-Schmidt procedure is
employed to build the basis states.

\vskip 6pt The remainder of this paper is organized as follows: in
section 2 the method used to derive the Hamiltonian is explained
in detail. In section 3, a brief overview of the model used for
the tubular arc is given. A periodic delta function potential is
also introduced to mimic atomic sites or defects. In section 4 the
method used to solve the Hamiltonian derived in section 2 is
briefly described. Results for the eigenvectors and eigenvalues of
the arc without delta functions, as well as the lowest energy
eigenvalues with delta sites, are presented. The conclusions are
given in section 5.\vskip 6pt \noindent $\bf{2. \ Derivation \ of
\ the \ Hamiltonian}$ \vskip 6pt Consider a quantum particle in
the neighborhood of a two-dimensional manifold $\Sigma $ imbedded
in$\ R^{3} $. Any point in the neighborhood of $\Sigma $ can be
given by the Monge form  plus a normal term
$$
 \vec r(q_1,q_2,q_3) = \vec x(q_1,q_2)+q_3 \hat e_3 \eqno (1)
$$
where $\vec x $ describes the manifold, and $\hat e_3 $ is the
unit vector normal to the surface. Applying the exterior
derivative operator [21] $\ d $ gives the one-form
$$
 d\vec r = d\vec x +dq_3 \hat e_3 +q_3 d\hat e_3 \eqno
(2)
$$
where$\ dq_3 $ is the incremental displacement along the normal to
the surface.$\ d\vec x $ lies in the tangent plane and is given by
$$
 d\vec x = \sigma_1 \hat e_1 +\sigma_2 \hat e_2 \eqno
(3)
$$
$\hat e_1 $ and $\hat e_2 $ are locally orthonormal unit vectors
tangent to the surface; $\sigma_1 $ and $\sigma_2 $ are one-forms
on $\Sigma $. The exterior derivative of any zero-form (scalar
function) $\psi $ in the neighborhood of $\Sigma $ is given by
$$
 d\psi = {\partial\psi \over \partial q_1}dq_1
+{\partial\psi \over \partial q_2}dq_2 +{\partial\psi \over
\partial q_3}dq_3
=\nu_1 \tau_1 +\nu_2 \tau_2+\nu_3 dq_3  \eqno (4)
$$
$\nu_1 $, $\nu_2 $, and $\nu_3 $ are zero-form functions and
$\tau_1 $, $\tau_2 $, and$\ dq_3 $ are the one-forms in the
neighborhood of the surface $\Sigma $.\vskip 6pt Applying the
Hodge * operator to the one-form of Eq. (4) gives
$$
 *d\psi = \nu_1 \tau_2 dq_3 +\nu_2 dq_3 \tau_1 +\nu_3
\tau_1 \tau_2 \eqno (5)
$$
A second application of the exterior derivative operator yields
$$
 d*d\psi = (\triangle\psi)\tau_1 \tau_2 dq_3 \eqno (6)
$$
where $\triangle $ is the Laplace-Beltrami operator. The time
independent Schrodinger equation becomes
$$
 -{\hbar^{2}\over 2m^{*}}\triangle\psi(\vec r)+V(\vec
r)\psi(\vec r) = E\psi(\vec r) \eqno (7)
$$
\vskip 6pt In the development above, no constraint has been
imposed on the particle. To bring the particle to the surface, we
imagine an oscillator potential$\ V(q_3)={1 \over
2}m^{*}\varpi^{2}q_3^{2} $ everywhere normal to $\Sigma $ acting
to enforce the condition that the particle remains in the
neighborhood of $\Sigma $. In the limit that
$\varpi\rightarrow\infty $, then$\ q_3\rightarrow 0 $. However, it
is not enough simply to include this term in Eq. (7); to preserve
the norm as the particle approaches the surface, it must further
hold that for$\ q_3\rightarrow 0 $ [2,3,4]
$$
 |\psi(q_1,q_2,q_3)|^{2}FdSdq_3\rightarrow
|\chi(q_1,q_2,q_3)|^{2}dSdq_3 \eqno(8)
$$
which allows the identification
$$
 \psi = {\chi\over \sqrt{F}} \eqno (9)
$$
where
$$
 F=1+2q_3H+q_3^{2}K
$$
$\ H $ and$\ K $ are the mean and Gauss curvatures respectively and
only depend on the surface coordinates$\ q_1 $ and $\ q_2 $.
Performing the standard factorization of $\chi $ into tangential and
normal parts and taking the$\ q_3\rightarrow 0 $ limit gives two
equations when Eq. (9) is substituted into Eq. (7),
$$
 -{\hbar^{2}\over 2m^{*}}\triangle_t\Psi -
{\hbar^{2}\over 2m^{*}}(H^{2}-K)\Psi =\varepsilon\Psi \eqno (10)
$$
and
$$
 -{\hbar^{2}\over 2m^{*}}{\partial^{2}\Phi\over
\partial q_3^{2}}+V(q_3)\Phi = \epsilon\Phi \eqno (11)
$$
The derivative free term in Eq. (10) is the distortion potential$\
V_D $, dependent on the mean curvature$\ H $ and the Gauss
curvature$\ K $ of the surface
$$
 V_D(q_1,q_2) = -{\hbar^{2}\over 2m^{*}}(H^{2}-K) \eqno
(12)
$$
\vskip 6pt\noindent It should be noted that$\ V_D $ is not
necessarily the only modification to the Laplace-Beltrami
operator. Even for surfaces possessing symmetry, factors related
to the metric can appear in the kinetic-energy operator and in
general cannot be transformed away [22]. \vskip 6pt To apply the
formalism described above to nanotubes, consider a tube of
radius$\ a $ and let $ u $ be the coordinate axis along $\hat i $.
The surface of the tube can be described by
$$
 \vec x(\theta,u) = (u-a\beta\cos{\theta})\hat i
+(f(u)+a\alpha\cos{\theta})\hat j +(a\sin{\theta})\hat k \eqno
(13)
$$
where$\ f(u) $ is the shape function for the axis of the tube,
$\alpha=(1+f_u^{2})^{-1/2} $, $\beta=(1+f_u^{2})^{-1/2}f_u $, and
$\ f_u=\partial_u f(u) $. Applying $\ d $ to Eq. (13) and using
Eq. (3), we find that
$$
 \sigma_1 =a d\theta \eqno(14a)$$ $$ \sigma_2 =\lambda(\theta,s) ds \eqno
(14b)
$$
and
$$
\hat e_1 =\beta(s)\sin{\theta} \ \hat i-\alpha(s)\sin{\theta} \ \hat
j+\cos{\theta} \ \hat k \eqno(15a)$$
$$ \hat e_2 =\alpha(s)\hat i +\beta(s)\hat j  \eqno(15b)$$ $$ \hat e_3 =
-\beta(s)\cos{\theta} \ \hat i +\alpha(s)\cos{\theta} \ \hat j
+\sin{\theta} \ \hat k \eqno (15c)
$$
with the integration measure, axis curvature, and arclength given
by $\lambda(\theta,s)=1-a\kappa(s)\cos{\theta} $, $\kappa
=(1+f_u^{2})^{-3/2}f_{uu} $, and $ ds=\sqrt{1+f_u^{2}}du $,
respectively. Any point in the neighborhood of the tube is given
by Eq. (1). The Laplace-Beltrami operator can thus be written
$$
 \triangle
={1\over\mu^{2}}({\partial^{2}\over\partial\theta^{2}}
+{\partial\ln\Lambda\over\partial\theta}{\partial\over\partial\theta})
+{1\over\Lambda^{2}}({\partial^{2}\over\partial s^{2}}
-{\partial\ln\Lambda\over\partial s}{\partial\over\partial s})
+{\partial^{2}\over\partial q_3^{2}}
+{\partial\ln(\Lambda\mu)\over\partial q_3}{\partial\over\partial
q_3} \eqno (16)
$$
Here
$$
 \mu =a(1+q_3\kappa_1)\eqno(17a)$$ $$ \Lambda
=\lambda(1+q_3\kappa_2) \eqno (17b)
$$
where $\kappa_1 $ and $\kappa_2 $ are the principal curvatures of
the tube. The Gaussian curvature$\ K $ and mean curvature$\ H $ are
defined in terms of the principal curvatures by the relationships
$$
 K =\kappa_1\kappa_2 \eqno(18a)$$ $$ H ={\kappa_1+\kappa_2\over 2}
\eqno (18b)
$$
For our surface parameterizations the principal curvatures are
found to be
$$
 \kappa_1 ={1\over a} \eqno(19a)$$ $$ \kappa_2
=-{\kappa(s)\cos{\theta}\over{1-a\kappa(s)\cos{\theta}}} \eqno
(19b)
$$
In the limit$\ q_3\rightarrow 0 $, the kinetic energy operator
reduces to
$$
 \triangle_t ={1\over\
a^{2}}({\partial^{2}\over\partial\theta^{2}}
+{\partial\ln\lambda\over\partial\theta}{\partial\over\partial\theta})
+{1\over\lambda^{2}}({\partial^{2}\over\partial s^{2}}
-{\partial\ln\lambda\over\partial s}{\partial\over\partial s})
\eqno (20)
$$
with the distortion potential given by
$$
 V_D(\theta,s) =-{\hbar^{2}\over 8m^{*}a^{2}}{1\over
[1-a\kappa(s)\cos{\theta}]^{2}} \eqno (21)
$$
\vskip 6pt\noindent Thus, for a quantum particle constrained to move
on the surface of a tube, the surface Hamiltonian becomes
$$
H^{(c)}=-{\hbar^{2}\over 2m^{*}}\triangle_t+V_D(\theta,s) \eqno
(22)
$$
\vskip 6pt\noindent and
$$
H^{(n)}=-{\hbar^{2}\over 2m^{*}}{\partial^{2}\over
\partial q_3^{2}}+V(q_3) \eqno (23)
$$
\vskip 6pt\noindent is the Hamiltonian due to the normal term.

\vskip 6pt\noindent $\bf{3. \ The \ Model}$\vskip 6pt The distorted
parts of  nanostructures often occur over relatively small sections
of the object, so as a first step we choose to model a finite tube
with pronounced curvature only over a small region with hard wall
boundary conditions at each end. Further simplification follows if
we are able to explicitly parameterize by the arclength. As will be
seen in the next section, this will facilitate performing the large
number of integrations that must be carried out when computing the
Gram-Schmidt coefficients and the surface Hamiltonian matrix
elements. With this in mind, we choose as our shape function
$$
f(u)=-{1\over \kappa_0}\cosh{\kappa_0(u-u_0)} \eqno (24)
$$
\vskip 6pt\noindent where $\kappa_0 $ is the curvature parameter
and$\ u_0 $ is the turning point of the shape function along the$\ u
$ coordinate axis (see Figure 1). The shape function of Eq. (24) has
the advantage that the arclength as a function of $u$ is one-to-one
and analytically invertible. In terms of arclength, the curvature of
the axis is given by
$$
 \kappa(s)=-{\kappa_0\over 1+[\kappa_0 (s-s_0)]^2} \eqno (25)
$$
 \vskip 6pt\noindent where$\ s_0 $ is the turning point along the arclength. Thus $\kappa_0$ is the
magnitude of the curvature of the axis at the turning point.\vskip
6pt As a crude model of atomic sites of the nanotube or as model of
defects of a quantum waveguide, we add periodic $\delta$-function
site potentials to the Hamiltonian
$$
 V(\theta,s)=-\Lambda_0\sum_{j=1}^{N_{a}}{\sum_{k=1}^{N_{r}}{\delta(\theta-\theta_{jk})\delta(s-s_k)}}
\eqno (26)
$$
\vskip 6pt\noindent where $\Lambda_0 $ is the strength of the potential,
$\theta_{jk} $ is the value of the angle at the$\ j$th site on the$\ k
$th ring, and$\ s_k $ is the value of the arclength at the$\ k $th
ring.$\ N_{a} $ and$\ N_{r} $ are the number of angular sites per ring
and the number of rings, respectively.

\vskip 6pt\noindent$\bf{4. \ Solution \ Method \ and \ Results}$ \vskip
6pt There are many techniques at our disposal for arriving at the
eigenvectors and eigenvalues of curved nanostructures. In a recent work
[22], the eigenstates of a quantum particle constrained to move on the
surface of a torus were found using a power series technique. That
procedure could be adopted in [22] for two reasons,  both of which are
consequences of a high degree of symmetry. First, the distortion
potential was only dependent upon one coordinate. Secondly, the kinetic
energy operator was separable. However, those symmetries are not at play
here; the distortion potential is a function of both coordinates and the
kinetic energy operator is not separable. Thus, an alternative method
must be employed. The alternative chosen here is a basis set
expansion.\vskip 6pt The main difficulty faced when trying to expand
$\Psi$ in a complete set is that, to our knowledge, there is no known set
of functions $\{\varphi_k\}$ orthogonal over $\Sigma$, i.e., where
$$
 \int_\Sigma{\sigma_1\sigma_2\varphi_j^*(\theta,s)\varphi_k(\theta,s)=\delta_{jk}}
\eqno (27)
$$
\vskip 6 pt\noindent holds true. To overcome this difficulty, we
make use of the Gram-Schmidt procedure [23] extended to two
dimensions. We choose as our original basis the set of$\ (2M+1)N $
functions
$$
 \xi_j(\theta,s)\equiv\xi_{mn}(\theta,s)=e^{im\theta}\sin{n\pi
s\over L}
$$
\vskip 6pt\noindent where$\ L $ is the length of the tube,$\
m=-M,\ldots,-1,0,1,\ldots,M $,$\ n=1,\ldots,N$, and$\
|j=1\rangle\equiv |m=-M,n=1\rangle $,$\ |j=2\rangle\equiv
|m=-M+1,n=1\rangle $,$\ldots$, $\ |j=(2M+1)N\rangle\equiv
|m=M,n=N\rangle $. The orthonormal set $\{\varphi_j\}$ is
constructed from the non-orthogonal set $\{\xi_j\}$.\vskip 6pt Since
we have a legitimate basis, the algorithm is straightforward. The
Schrodinger equation is solved as a matrix eigenvector-eigenvalue
equation with the eigenvectors being the coefficients of the
expansion and the elements of the Hamiltonian matrix are defined by
$$
 H_{jk}^{(c)}=\int_\Sigma{\sigma_1\sigma_2\varphi_j^*(\theta,s)H^{(c)}\varphi_k(\theta,s)}
\eqno(28)
$$
\vskip 6pt\ Since the Hamiltonian in Eq. (22) is invariant under
$\theta \rightarrow -\theta$ we expect the solutions to segregate
themselves into positive and negative parity solutions. This will
prove useful as a check on the reliability of the numerical results.
It should also be noted that in the presence of delta functions, we
expect the number of angular peaks in the ground state probability
density to be equal to the number of delta sites per ring. This will
also serve as a useful check on the numerics.\vskip 6pt\ Using
$m^*=m_e$ the $\varepsilon<0$ results for the tubular arc without
delta function site potentials are presented in tables 1-4. As can
be seen, all $\varepsilon<0$, i.e. $m=0$ states, are of positive
parity. The $|m|>0$ states alternate between positive and negative
parity. Table 1 shows the results for the ground state of tubes with
several curvature parameters and turning points. Convergence was
achieved with 5-digit accuracy using a 20-state ($M=2,N=4$) basis
set expansion. When $\kappa_{0}=0$, the tube is straight and the
system can be solved analytically; the distortion potential is a
constant that can be subtracted from the eigenvalues but is left
here for comparison with the arc. It can be seen that the energy of
the ground state is lowered when the value of the curvature
parameter is increased. However, curvature has little effect on the
energies of the excited states (tables 2-4). It should also be noted
that the energy of the ground and excited states are sensitive to
the position of the bend as well. If the turning point is closer to
either end of the tube than it is to $ L/2$, the eigenenergies of
the ground, first and third excited states are again lowered,
however, the energy of the second excited state is raised.

\vskip 6pt Curvature and bend locations have an effect on the charge
(probability) density as well. Although the charge density of the
straight tube is angular independent, curvature induces angular
dependence in the charge density of the ground state that is maximum
at the point of minimum radius of curvature ($\theta=\pi,s=s_0$),
i.e. the point of maximum curvature. The charge density of the
ground state at $\theta=\pi$ is shown in Figure 2. There is no
angular dependence in the excited states, but curvature and bend
location still have an effect on the charge density. The nodes and
peaks of the density are shifted and the height of the peaks are no
longer even (Figs. 3-5).

\vskip 6pt The eigenvalues of the arc with 1170 ($N_{a}=6,N_{r}=195
$) delta function potentials arranged in an armchair configuration
are given in table 5. These results were obtained with a 52-state
($M=6,N=4$) expansion. The value for the strength of the potential
was chosen to be of the same order of magnitude as the deepest part
of the distortion potential well. As can be seen, the ground state
is more sensitive to the curvature than the excited states. The
slight increase in energy for the third excited states shown in the
table may be due to a need for more basis states in the expansion.
In the presence of the delta function potentials, charge
localization still occurs in the region of curvature (Figure 6);
however, there are several islands of localization, with the maximum
occurring near the center of the potential well. As expected, the
number of islands is equal to the number of delta function sites in
each ring, i.e., if there are 6 deltas per ring, there are 6 peaks.
\vskip 6pt\noindent $ \bf{5. \ Conclusion}$ \vskip 6pt The
eigenstates and eigenvalues of a particle constrained to move on the
surface of a tubular arc with and without delta function sites were
computed for several values of $\kappa_0$ and $s_0$. It was shown
that the energy and density were not only sensitive to the strength
of the curvature but also to the location of the bend. There was
strong interplay between the curvature and delta function potentials
demonstrated by the density plot of Figure 6.  Charge gets localized
by the deltas as well as curvature. The density is still peaked over
$s=s_0$, but the maximum peak is no longer at $\theta=\pi$, the
center of the well. This peak is slightly displaced due the deltas
competing with the curvature as well as each other. What this type
of interplay would mean for electron transport is an open question
and a topic currently under investigation. \vskip 6pt \noindent
\centerline{$\bf{Acknowledgements}$} \vskip 6pt L.M. would like to
acknowledge useful suggestions from Ms. Samanthia Long. L.M. and
M.E. received support from NASA Grant No. NAG2-1439. \vskip 6pt
\noindent \centerline{$\bf{References}$} \vskip 6pt \noindent 1. H.
Jensen and H. Koppe, Ann. of Phys. ${\bf 63}$, 586 (1971). \vskip
6pt

\noindent 2. R. C. T. da Costa,  Phys.  Rev. A {\bf 23} , 1982
(1981). \vskip 6pt

\noindent 3. R. C. T. da Costa,  Phys.  Rev. A {\bf 25} , 2893
(1982). \vskip 6pt

\noindent 4. P. Exner and P. Seba, J. Math. Phys. {\bf 30}, 2574
(1989). \vskip 6pt

\noindent
 5. P. Duclos and P. Exner,  Rev. in  Math.
Phys., ${\bf 7}$, 73 (1995). \vskip 6pt

\noindent 6. J.T. Londergan, J.P. Carini, D.P. Murdock, {\it Binding
and Scattering in Two Dimensional  Systems; Applications to Quantum
Wires, Waveguides, and Photonic crystals}, (Springer-Verlag Berlin,
1999). \vskip 6pt

\noindent 7. S. Matsutani, J. Phys. Soc. Jap.  {\bf 61}, 55 (1991).
\vskip 6pt

\noindent
 8. J. Goldstone and R. L. Jaffe, Phys. Rev. B ${\bf 45}$, 14100 (1991).
 \vskip 6pt

\noindent
 9. P. Ouyang, V. Mohta and R. L. Jaffe, Ann. of Phys. ${\bf 275}$,
297 (1998). \vskip 6pt

 \noindent
 10. S. Midgley, Aus. J. Phys ${\bf 53}$, 77 (2000).
 \vskip 6pt

\noindent
 11. S. Midgley and J.B. Wang,  Aus.  J.  Phys.,
{\bf 53}, 77 (2000). \vskip 6pt

\noindent 12. I. J. Clark and A. J. Bracken, J. Phys. A ${\bf 29}$,
4527 (1996).
 \vskip 6pt

\noindent 13. M. Encinosa and B. Etemadi,  Phys. Rev. A {\bf 58}, 77
(1998). \vskip 6pt

\noindent 14. M. Encinosa and B. Etemadi, Physica B ${\bf 266}$, 361
(1998).
 \vskip 6pt

\noindent 15. M. Encinosa,
  IEEE Trans.   Elec.  Dev., {\bf 47}, 878 (2000).\vskip 6pt

\noindent 16. R. Saito, G. Dresselhaus, and M.S. Dresselhaus, {\it
Physical Properties of Carbon Nanotubes}, (Imperial College Press,
London, 1998).\vskip 6pt

\noindent 17. H.R. Shea, R. Martel, and Ph. Avouris, Phys. Rev.
Lett. {\bf 84}, 4441 (2000). \vskip 6pt

\noindent 18. S. Matsutani, Phys. Rev. A {\bf 91}, 686 (1993).
\vskip 6pt

\noindent 19. S. Matsutani, J. Phys. A {\bf 30}, 4019  (1997).
\vskip 6pt

\noindent
 20. M. Burgess and B. Jensen, Phys. Rev. A {\bf 48}, 1861 (1993).
\vskip 6pt

\noindent 21. H. Flanders, {\it Differential Forms with Applications
to the Physical Sciences}, (Dover Publications, 1989).\vskip 6pt

\noindent 22. M. Encinosa and L. Mott, Phys. Rev. A {\bf 58}, 014102
(2003). \vskip 6pt

\noindent 23. G. Arfken and H. Weber, {\it Mathematical  Methods for
Physicists}, 5th ed., (Academic Press, New York, 1995). \vskip 6pt

\vfill \eject

\begin{table}
\caption{Ground state eigenfunctions and eigenvalues for the tubular
arc without delta function site potentials. For all tubes considered
here $a=0.85nm, L=100nm$. Coefficients not listed are at least an
order of magnitude smaller than those given. }
\begin{center}
\begin{tabular}{|ll|l|l|}
\hline
\ $\kappa_0$ & $\ \ s_0$ &\ \qquad  \qquad \qquad  \qquad  $\Psi_{mn}; a=0.85nm, L=100nm$ & $\ \ \ \varepsilon (meV)$\\
\hline
0.00 & ----- & $\Psi_{01}=.0612 \sin({\pi s \over 100})$ &-13.1423 \\
0.75 & 51.87 & $\Psi_{01}=-.0022\cos\theta \sin({\pi s\over 100})+.0555\sin({\pi s\over 100})-.0042\sin({\pi s\over 50})$ &-13.4068 \\
 &  & $+.0021\cos\theta \sin({3\pi s\over 100})-.0253\sin({3\pi s\over 100})+.0040\sin({\pi s\over 25})$& \\
0.75 & 55.60 & $\Psi_{01}=.0023\cos\theta \sin({\pi s\over 100})-.0541\sin({\pi s\over 100})-.0135\sin({\pi s\over 50})$ &-13.4262 \\
&  & $-.0019\cos\theta \sin({3\pi s\over 100})+.0227\sin({3\pi s\over 100})+.0014\cos\theta \sin({\pi s \over 25})-.0117\sin({\pi s\over 25})$& \\
0.95 & 52.37 & $\Psi_{01}=.0026\cos\theta \sin({\pi s\over 100})-.0526\sin({\pi s\over 100})+.0061\sin({\pi s\over 50})$ &-13.5793 \\
&  & $-.0024\cos\theta \sin({3\pi s\over 100})+.0302\sin({3\pi s\over 100})-.0066\sin({\pi s\over 25})$& \\
0.95 & 57.08 & $\Psi_{01}=-.0027\cos\theta \sin({\pi s\over 100})+.0500\sin({\pi s\over 100})+.0012\cos\theta \sin({\pi s\over 50})$ &-13.6313\\
&  & $-.0179\sin({\pi s\over 50})+.0021\cos\theta \sin({3\pi s\over 100})-.0248\sin({3\pi s\over 100})-.0020\cos\theta \sin({\pi s\over 25})$& \\
&  & $+.0183\sin({\pi s\over 25})$&\\
1.00 & 52.50 & $\Psi_{01}=.0027\cos\theta \sin({\pi s\over 100})-.0517\sin({\pi s\over 100})+.0065\sin({\pi s\over 50})$ &-13.6522  \\
&  & $-.0025\cos\theta \sin({3\pi s\over 100})+.0315\sin({3\pi s\over 100})-.0074\sin({\pi s\over 25})$& \\
1.00 & 57.45 & $\Psi_{01}=-.0028\cos\theta \sin({\pi s\over 100})+.0488\sin({\pi s\over 100})+.0013\cos\theta \sin({\pi s\over 50})$ &-13.7191 \\
&  & $-.0189\sin({\pi s\over 50})+.0021\cos\theta \sin({3\pi s\over 100})-.0250\sin({3\pi s\over 100})-.0022\cos\theta \sin({\pi s\over 25})$& \\
&  & $+.0202\sin({\pi s\over 25})$&\\
1.15 & 73.79 & $\Psi_{02}=.0027\cos\theta \sin({\pi s\over 100})-.0356\sin({\pi s\over 100})-.0036\cos\theta \sin({\pi s\over 50})$ &-14.2313 \\
&  & $+.0438\sin({\pi s\over 50})+.0022\cos\theta \sin({3\pi s\over 100})-.0235\sin({3\pi s\over 100})-.0048\sin({\pi s\over 25})$& \\
\hline
\end{tabular}
\end{center}
\end{table}
\begin{table}
\caption{Eigenfunctions and eigenvalues of the first excited states
of the tubular arc. Coefficients not listed are at least an order of
magnitude smaller than those given. }
\begin{center}
\begin{tabular}{|ll|l|l|}
\hline
\ $\kappa_0$ & $\ \ s_0$ &\ \qquad  \qquad \qquad  \qquad  $\Psi_{mn}; a=0.85nm, L=100nm$ & $\ \ \ \varepsilon (meV)$\\
\hline
0.00 & ----- & $\Psi_{02}=.0612\sin({\pi s \over 50})$ &-13.0296 \\
0.75 & 51.87 & $\Psi_{02}=.0067\sin({\pi s\over 100})+.0606\sin({\pi s\over 50})+.0045\sin({3\pi s\over 100})$ &-13.0308 \\
0.75 & 55.60 & $\Psi_{02}=.0185\sin({\pi s\over 100})+.0577\sin({\pi s\over 50})+.0084\sin({3\pi s\over 100})-.0025\sin({\pi s\over 25})$ &-13.043 \\
0.95 & 52.37 & $\Psi_{02}=.0103\sin({\pi s\over 100})+.0600\sin({\pi s\over 50})+.0061\sin({3\pi s\over 100})$ &-13.0308 \\
0.95 & 57.08 & $\Psi_{02}=.0257\sin({\pi s\over 100})+.0545\sin({\pi s\over 50})+.0098\sin({3\pi s\over 100})-.0038\sin({\pi s\over 25})$ & -13.0476\\
1.00 & 52.50 & $\Psi_{02}=-.0114\sin({\pi s\over 100})-.0597\sin({\pi s\over 50})-.0067\sin({3\pi s\over 100})-.0011\sin({\pi s\over 25})$ & -13.0312  \\
1.00 & 57.45 & $\Psi_{02}=-.0275\sin({\pi s\over 100})-.0536\sin({\pi s\over 50})-.0100\sin({3\pi s\over 100})+.0041\sin({\pi s\over 25})$ &-13.0497 \\
1.15 & 73.79 & $\Psi_{01}=.0488\sin({\pi s\over 100})+.0365\sin({\pi s\over 50})-.0057\sin({3\pi s\over 100})-.0012\sin({\pi s\over 25})$ & -13.1016 \\
\hline
\end{tabular}
\end{center}
\end{table}
\begin{table}
\caption{Eigenfunctions and eigenvalues of the second excited states
of the tubular arc. Coefficients not listed are at least an order of
magnitude smaller than those given. }
\begin{center}
\begin{tabular}{|ll|l|l|}
\hline
\ $\kappa_0$ & $\ \ s_0$ &\ \qquad  \qquad \qquad  \qquad  $\Psi_{mn}; a=0.85nm, L=100nm$ & $\ \ \ \varepsilon (meV)$\\
\hline
0.00 & ----- & $\Psi_{03}=.0612\sin({3\pi s \over 100})$ &-12.8416 \\
0.75 & 51.87 & $\Psi_{03}=.0249\sin({\pi s\over 100})-.0068\sin({\pi s\over 50})+.0554\sin({3\pi s\over 100})-.0035\sin({\pi s\over 25})$ &-12.9331 \\
0.75 & 55.60 & $\Psi_{03}=-.0211\sin({\pi s\over 100})+.0151\sin({\pi s\over 50})-.0546\sin({3\pi s\over 100})+.0093\sin({\pi s\over 25})$ &-12.9169 \\
0.95 & 52.37 & $\Psi_{03}=-.0295\sin({\pi s\over 100})+.0103\sin({\pi s\over 50})-.0524\sin({3\pi s\over 100})+.0049\sin({\pi s\over 25})$ &-12.9478 \\
0.95 & 57.08 & $\Psi_{03}=-.0225\sin({\pi s\over 100})+.0208\sin({\pi s\over 50})-.0516\sin({3\pi s\over 100})+.0012\sin({\pi s\over 25})$ & -12.9182\\
1.00 & 52.50 & $\Psi_{03}=.0306\sin({\pi s\over 100})-.0115\sin({\pi s\over 50})+.0514\sin({3\pi s\over 100})-.0053\sin({\pi s\over 25})$ & -12.9515  \\
1.00 & 57.45 & $\Psi_{03}=-.0227\sin({\pi s\over 100})+.0221\sin({\pi s\over 50})-.0508\sin({3\pi s\over 100})+.0013\sin({\pi s\over 25})$ & -12.9178 \\
1.15 & 73.79 & $\Psi_{03}=-.0100\sin({\pi s\over 100})+.0222\sin({\pi s\over 50})+.0559\sin({3\pi s\over 100})+.0041\sin({\pi s\over 25})$ & -12.874 \\
\hline
\end{tabular}
\end{center}
\end{table}
\begin{table}
\caption{Eigenfunctions and eigenvalues of the third excited states
of the tubular arc. Coefficients not listed are at least an order of
magnitude smaller than those given. }
\begin{center}
\begin{tabular}{|ll|l|l|}
\hline
\ $\kappa_0$ & $\ \ s_0$ &\ \qquad  \qquad \qquad  \qquad  $\Psi_{mn}; a=0.85nm, L=100nm$ & $\ \ \ \varepsilon (meV)$\\
\hline
0.00 & ----- & $\Psi_{04}=.0612\sin({\pi s \over 25})$ &-12.5784 \\
0.75 & 51.87 & $\Psi_{04}=-.0022\sin({\pi s\over 100})+.0048\sin({3\pi s\over 100})+.0610\sin({\pi s\over 25})$ &-12.5799 \\
0.75 & 55.60 & $\Psi_{04}=.0065\sin({\pi s\over 100})-.0027\sin({\pi s\over 50})-.0133\sin({3\pi s\over 100})-.0593\sin({\pi s\over 25})$ &-12.6136 \\
0.95 & 52.37 & $\Psi_{04}=.0034\sin({\pi s\over 100})-.0074\sin({3\pi s\over 100})-.0606\sin({\pi s\over 25})$ &-12.5805 \\
0.95 & 57.08 & $\Psi_{04}=.0093\sin({\pi s\over 100})-.0049\sin({\pi s\over 50})-.0194\sin({3\pi s\over 100})-.0571\sin({\pi s\over 25})$ & -12.6398\\
1.00 & 52.50 & $\Psi_{04}=.0038\sin({\pi s\over 100})+.0013\sin({\pi s\over 50})-.0083\sin({3\pi s\over 100})-.0605\sin({\pi s\over 25})$ & -12.5804  \\
1.00 & 57.45 & $\Psi_{04}=.0102\sin({\pi s\over 100})-.0056\sin({\pi s\over 50})-.0211\sin({3\pi s\over 100})-.0562\sin({\pi s\over 25})$ & -12.6475 \\
1.15 & 73.79 & $\Psi_{04}=.0012\sin({\pi s\over 100})-.0028\sin({\pi s\over 50})+.0057\sin({3\pi s\over 100})-.0609\sin({\pi s\over 25})$ & -12.25592 \\
\hline
\end{tabular}
\end{center}
\end{table}
\begin{table}
\caption{Eigenvalues of the tubular arc with 1170
($N_{a}=6,N_{r}=195 $) $\delta$-function site potentials arranged in
an armchair configuration with $\Lambda_{0}=400 meV \cdot nm$. The
subscripts on the eigenvalues are \textit{not} quantum numbers. They
refer to the order of the energies: 0 is the ground state; 1 is the
first excited state and so on.}
\begin{center}
\begin{tabular}{|ll|l|l|l|l|}
\hline
\ $\kappa_0$ & $\ \ s_0$ & $\ \ \ \varepsilon_{0} (meV)$ & $\ \ \ \varepsilon_{1} (meV)$ & $\ \ \ \varepsilon_{2} (meV)$ & $\ \ \ \varepsilon_{3} (meV)$\\
\hline
0.00 & --- & -872.208 & -872.095 & -871.907 & -871.643\\
0.95 & 52.37 & -872.587 & -872.115 & -872.012 & -871.627\\
1.00 & 52.50 & -872.655 & -872.119 & -872.018 & -871.624\\
\hline
\end{tabular}
\end{center}
\end{table}
\begin{figure}
\begin{center}
\leavevmode \epsffile{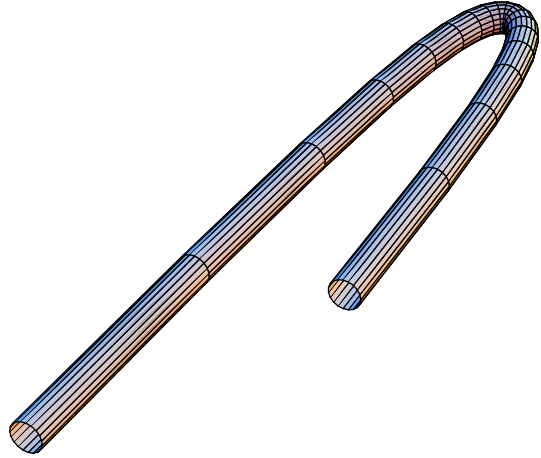}
\end{center}
\caption{A tubular arc.}
\end{figure}
\begin{figure}
\begin{center}
\leavevmode \epsffile{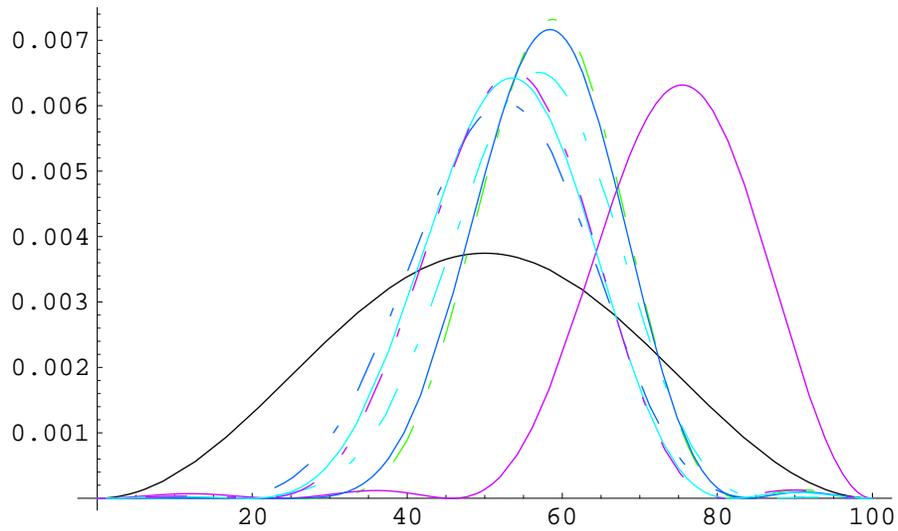}
\end{center}
\caption{Probability densities of the states given in Table I at
$\theta=\pi$. The color code is as follows and is the same for the
remaining figures: solid black ($\kappa_{0}=0.00$); dashed blue
($\kappa_{0}=0.75, s_{0}=51.87$); dashed turquoise
($\kappa_{0}=0.75, s_{0}=55.60$); solid turquoise ($\kappa_{0}=0.95,
s_{0}=52.37$); solid blue ($\kappa_{0}=0.95, s_{0}=57.08$); dashed
pink ($\kappa_{0}=1.00, s_{0}=52.50$); dashed green
($\kappa_{0}=1.00, s_{0}=57.45$); solid pink ($\kappa_{0}=1.15,
s_{0}=73.79$).}
\end{figure}
\begin{figure}
\begin{center}
\leavevmode \epsffile{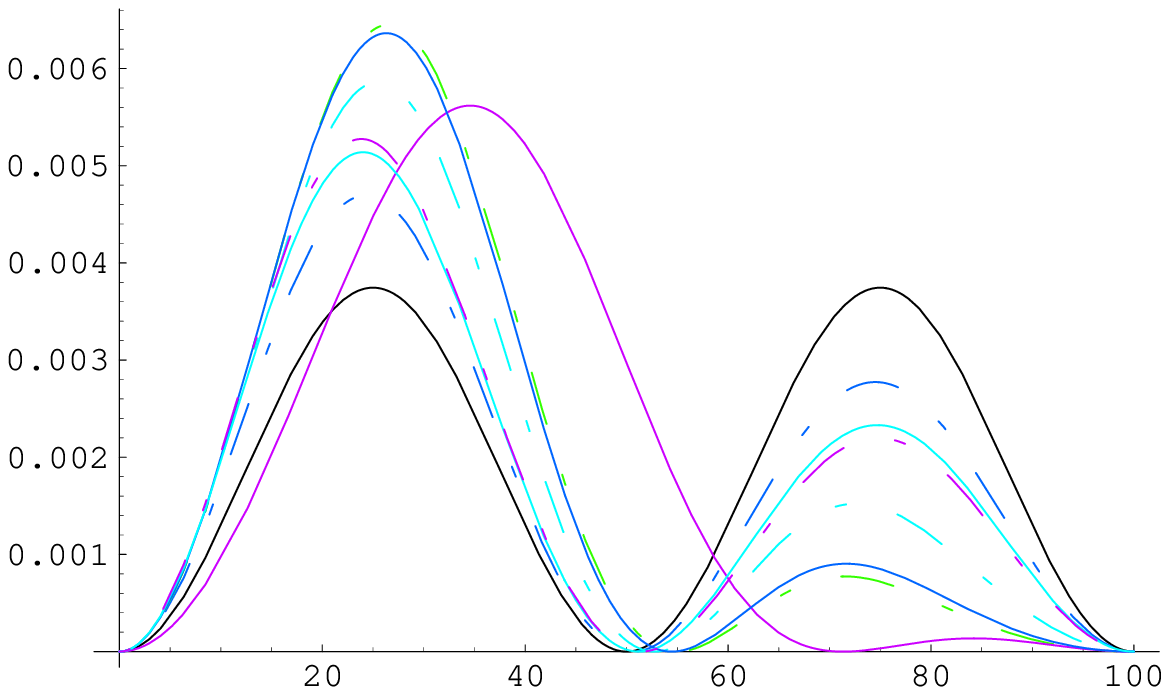}
\end{center}
\caption{Probability densities of the states given in Table II at
$\theta=\pi$.}
\end{figure}
\begin{figure}
\begin{center}
\leavevmode \epsffile{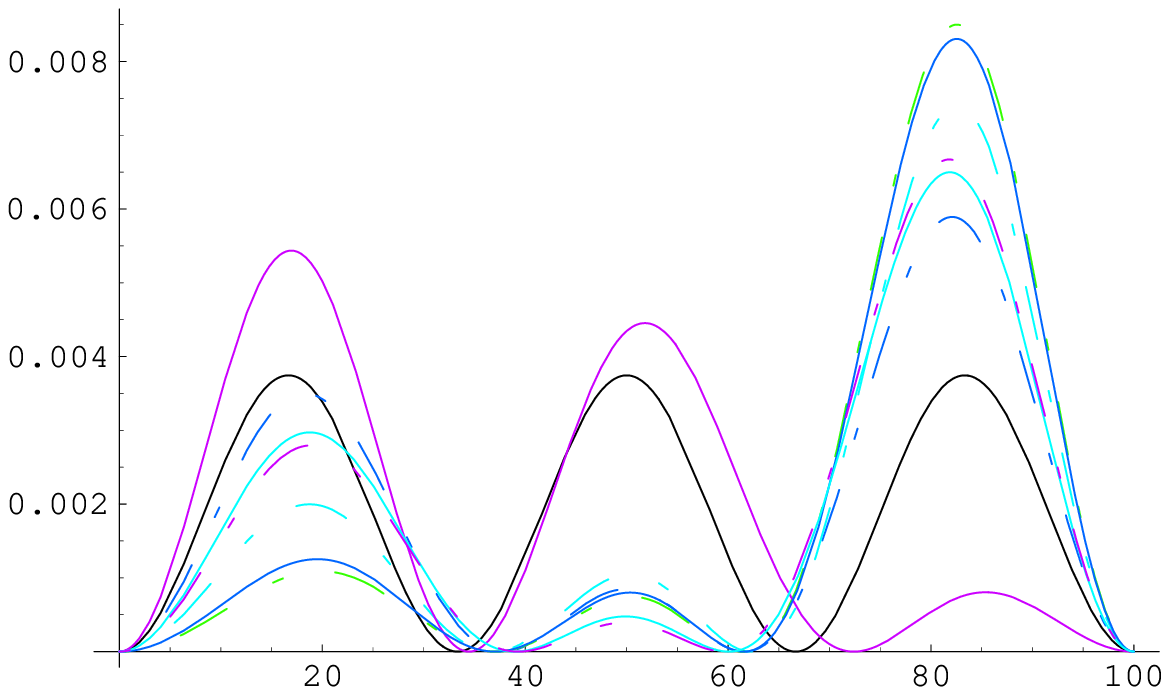}
\end{center}
\caption{Probability densities of the states given in Table III at
$\theta=\pi$.}
\end{figure}
\begin{figure}
\begin{center}
\leavevmode \epsffile{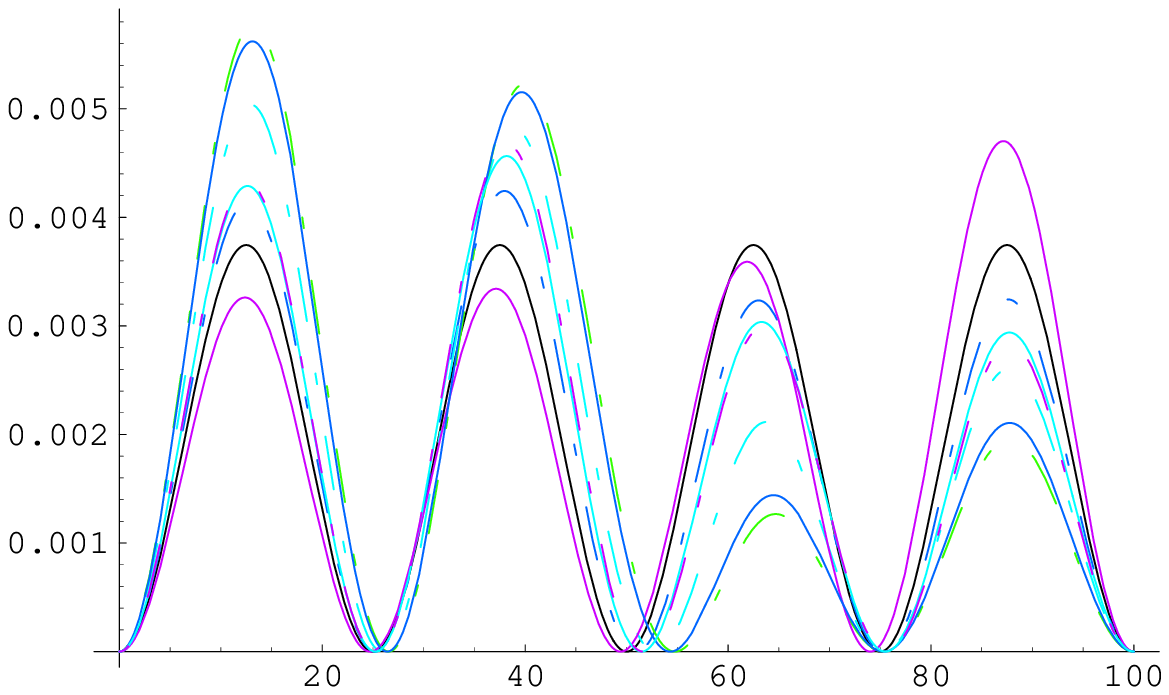}
\end{center}
\caption{Probability densities of the states given in Table IV at
$\theta=\pi$.}
\end{figure}
\begin{figure}
\begin{center}
\leavevmode \epsffile{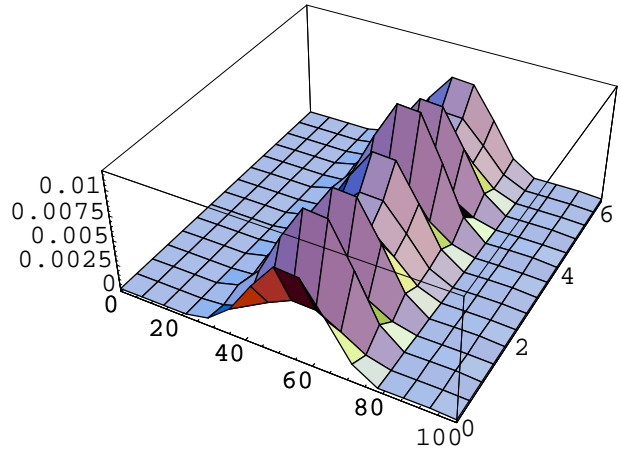}
\end{center}
\caption{Probability density of the ground state of a tubular arc
($\kappa_{0}=1.00, s_{0}=57.45$) with delta function site
potentials.}
\end{figure}

\end{document}